# Limb Projections on Saturn

Jeremy Shears and Richard Baum

## Abstract

We present two unusual observations of Saturn recorded during the early years of the twentieth century

## Text

Whilst carrying out research into the life of Thomas Hinsley Astbury (1858-1922; Figure 1), one of the authors (JS) came across an unusual observation he made of Saturn during the lunar occultation on 1900 September 3, which might be of interest to readers, especially planetary observers [1]. The eagle-eyed Astbury, best known for his variable star discoveries, was observing from his home in Wallingford, Oxon, using his 3¼ inch (8.3 cm) Wray refractor x80. Conditions were pretty much as good as they could be considering that the planet was in Sagittarius, thus at a low elevation, and at the time of the observation he noted 'definition was perfect'. As he peered at the planet as the occultation proceeded, he could see the Cassini division and the shadow of the planet on the rings quite clearly and steadily, but he was not prepared for what he saw next. In his own words from his report published in the BAA *Journal*: [2]

'The first indication of anything unusual was a very small, but distinct, prominence just round the bend of the W. ansae towards the S. edge of the ring, sharply defined and of the same brilliance itself. A steady look convinced me of its reality, and I watched as it quickly decreased and vanished, just as the moon's dark limb was seen encroaching upon the outer ring, having lasted almost four or five seconds. It was situated some distance from the point of first contact, and its altitude was rather less that the distance across the outer ring at the point where it appeared.'

He went on to note that 'It is, of course, very easy to dismiss the question as an optical illusion; but I have never witnessed anything of the kind before, though I have observed Saturn some scores of times with the same instrument.'

Four years later, and still perplexed by his observation, he penned a further letter to the *Journal* [3]. 'I have just been reading, ' he wrote, 'that in 1876 Prof. Newcomb watched the occultation of Saturn' for possible effects indicative of a thin lunar atmosphere. Could it thus be, Astbury wondered, that differential refraction by this atmosphere as Saturn approached the limb of the moon, the phenomenon sought by Newcomb, was the explanation for what he had seen in 1900? 'At the point where the "flattened" portion of the ring met the undistorted part,' Astbury argued, 'the different curvatures would produce an angle or protuberance, and it was doubtless the upper, *i.e.,* southern, of these which I saw and described as a prominence. Being altogether unprepared, and not having considered the form which distortion might be expected to assume, I did not look for the northern "prominence"; but this would probably have been much less conspicuous, owing to the obliquity with which the ring system was occulted by the moon's limb.'

Of course we now know the moon's atmosphere is so tenuous that it could not cause such effects. So the question remains: what did Astbury observe? Was it a real effect, or simply an illusion? A. F. O'D. Alexander's classic work *The Planet Saturn - A History of Observation, Theory and Discovery* makes no reference to the observation, but does refer to a phenomenon of similar characteristics reported by the twenty year old Italian astronomer Mentore Maggini (1890-1941; Figure 2). [4]

Maggini, then an assistant at the Osservatorio Ximeniano, Florence, a small observatory established in 1756 by the Jesuit Leonardo Ximenes (1716-1786), announced his observation in the *Astronomisches Nachrichten* of 1910 October 11 [5] and published a full account in *Bulletin de la Société Astronomique de France* the following March [6].

Observing Saturn on 1910 September 29 with a Calver reflector of 35-cm aperture x350 he noticed at the extremity of the south equatorial belt, a large bright area, and close by, highlighted by the planet's shadow on the rings, a conspicuous luminous spot projecting from the west limb of the planet (Figure 3). No moon was in the vicinity at the time of the observation. The observation has never been fully explained.

"Irradiating spots, diagonal wisps and straight streaks," commented Alexander, "are more familiar features on Jupiter than on Saturn; the bright projection may have been a contrast effect caused by the contiguity of the bright spot and the dark shadow." [7]

A plausible if predictable supposition amply supported by the observational record of Venus and Mars. The eye is easily misled, and inference too readily subject to imagination. Even so other explanations cannot be excluded. The recent observations of impacts on Jupiter [8] have taught us to be more circumspect. Imperfect as the historical record is, in hindsight its anomalies occasionally prove surprisingly relevant and provide useful insights towards a better understanding of past effort, and present results. Importantly they remind us that (a) our predecessors were less fortunate in their equipment than ourselves and (b) that resolution is an evolutionary process dependent on technological progress, and on observers' individual visual experience, the visual experience of their milieux, and expectations formed by previous visual models.

**References**

[1] A detailed biography of T.H. Astbury will appear in a future edition of this *Journal*: Shears J., JBAA, accepted for publication (2012)
[2] Astbury T.H., JBAA, **11**, 35-36 (1900)
[3] Astbury T.H., JBAA, **15**, 40 (1904). The source of Astbury's continued interest is to be found in D P Todd, AN **90** (No. 2146), 159-160 (1877). Simon Newcomb (1835 – 1909) was primarily a theoretical astronomer. Not widely known is the fact he published *His Wisdom The Defender* (1900), a narrative that comes under the heading of futuristic fiction, in which he promulgates the use of aircraft to end war.
[4] Alexander A. F. O'D. *The Planet Saturn: A History of Observation, Theory and Discovery*, (London: Faber and Faber 1962), 323-324
[5] Maggini M., AN, **186**, 79 (1910)
[6] Maggini M., BSAF, March (1911)
[7] Alexander, op.cit. [4], 324.
[8] Some 15 years after the bombardment of Jupiter by comet Shoemaker-Levy 9, which caused visible impact scars, a new impact scar appeared on 2009 July 19, discovered by Anthony Wesley from New South Wales, Australia. See: Rogers J., JBAA, **119**, 235 (2009)

**Acknowledgements**

We are indebted to Tracey Wernham Clark & Emma Anderson for permission to reproduce the photograph of Astbury from their book *St. John's County Primary School Wallingford - Celebrating One Hundred Years of an Oxfordshire Market Town School*, Donatella Randazzo, Librarian at the INAF-Osservatorio Astronomico di Palermo, Italy, for providing the portrait of Maggini and Julian Baum for preparing the drawing of Saturn for publication. We thank our referees, Mike Frost and Mike Foulkes, Directors of the BAA Historical and Saturn Sections respectively, for their comments.

**Addresses**

JS: "Pemberton", School Lane, Bunbury, Tarporley, Cheshire, CW6 9NR, UK [bunburyobservatory@hotmail.com]

RB: 25 Whitchurch Road, Chester, CH3 5QA, UK [richard@take27.co.uk]

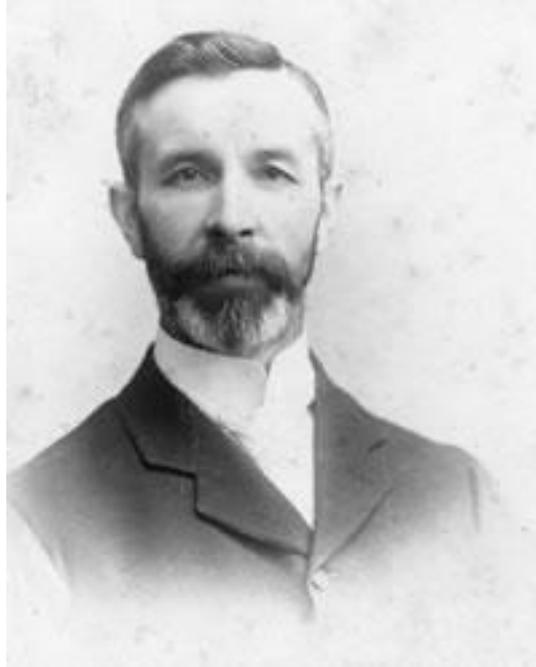

**Figure 1: Thomas Hinsley Astbury (1858-1922)**
Courtesy of Tracey Wernham Clark & Emma Anderson

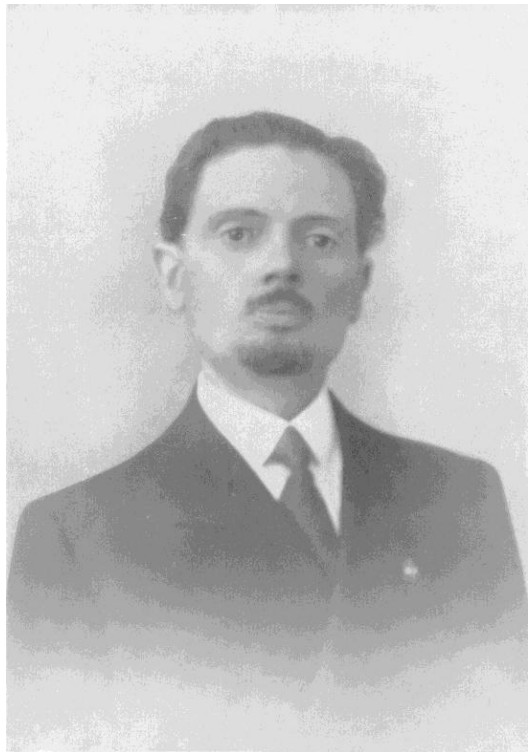

**Figure 2: Mentore Maggini (1890-1941)**
Courtesy of Library of INAF-Osservatorio Astronomico di Palermo, Italy

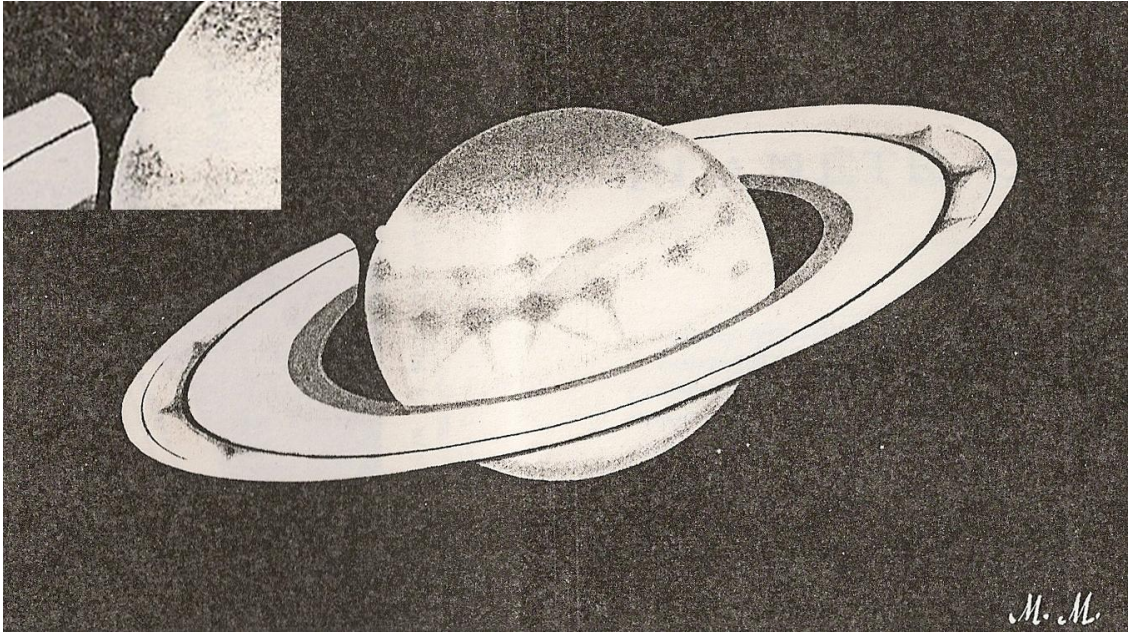

**Figure 3: Maggini's drawing of Saturn on 1910 September 29**

The inset shows an expanded view of the region around the spot projecting from the west limb of the planet